\documentclass[a4paper]{article}

\usepackage{INTERSPEECH2022}
\usepackage{booktabs}
\usepackage{multirow}
\newcommand{\share}[1]{\langle{#1}\rangle}

\title{Towards End-to-End Private Automatic Speaker Recognition}
\name{Francisco Teixeira$^1$, Alberto Abad$^1$, Bhiksha Raj$^2$, Isabel Trancoso$^1$}
\address{
  $^1$INESC-ID/Instituto Superior Técnico, University of Lisbon, Portugal\\
  $^2$LTI, Carnegie Mellon University, USA
\thanks{This work was supported by Portuguese national funds through Fundação para a Ciência e a Tecnologia (FCT), with references UIDB/50021/2020 and CMU/TIC/0069/2019.}}
\email{francisco.s.teixeira@tecnico.ulisboa.pt}

\begin{document}

\maketitle

\begin{abstract}
The development of privacy-preserving automatic speaker verification systems has been the focus of a number of studies with the intent of allowing users to authenticate themselves without risking the privacy of their voice. However, current privacy-preserving methods assume that the template voice representations (or speaker embeddings) used for authentication are extracted locally by the user. 
This poses two important issues: first, knowledge of the speaker embedding extraction model may create security and robustness liabilities for the authentication system, as this knowledge might help attackers in crafting adversarial examples able to mislead the system; second, from the point of view of a service provider the speaker embedding extraction model is arguably one of the most valuable components in the system and, as such, disclosing it would be highly undesirable.
In this work, we show how speaker embeddings can be extracted while keeping both the speaker's voice and the service provider's model private, using Secure Multiparty Computation. 
Further, we show that it is possible to obtain reasonable trade-offs between security and computational cost.
This work is complementary to those showing how authentication may be performed privately, and thus can be considered as another step towards fully private automatic speaker recognition.
\end{abstract}


\section{Introduction}
\label{sec:intro}

Recent years have seen an increase in the number of online services and applications that use speech as a means for authentication and interaction. 
Among other speech technologies, voice-based authentication systems -- or Automatic Speaker Verification (ASV) systems -- are becoming more and more a part of our everyday lives. The uniqueness and ubiquitous nature of speech make its use a straightforward manner to protect and grant access to both local and remote systems. 
However, in the remote case, ASV systems raise multiple privacy concerns. This comes from the fact that speech and the information it carries are extremely sensitive in nature: from speech one can extract information as diverse as the speaker's age, gender, emotional or personality traits, health state, among many others \cite{singh2019profiling, nautsch2019preserving}. As such, by sending their voice -- or a template thereof -- to a remove server, users are risking their privacy.
These concerns are reflected in recent regulations, as shown, for instance, by the definition of personal data provided by the European Union's General Data Protection Regulation (GDPR) \cite{gdpr}, wherein speech data and the information extracted from it are considered as Personally Identifiable Information, i.e., information that, on its own, allows determining the identity of an individual, and that, as such, is protected under the GDPR \cite{nautsch2019gdpr}. 

Due to the above, the problem of protecting privacy in the setting of ASV has been a precursor for much of the research done for speech privacy.
One of the first strides in this direction is the cryptographic-based work of Pathak et al. \cite{pathak2013privacy}, who adapted a Gaussian Mixture Model (GMM) to work with Homomorphic Encryption (HE) in order to perform speaker verification. Similarly, Port\^{e}lo et al. \cite{portelo2014privacy} implemented a privacy-preserving GMM-based speaker verification using Garbled Circuits. More recently, Nautsch et al.~and Treiber et al.~applied Homomorphic Encryption \cite{nautsch2018homomorphic} and Secure Multiparty Computation \cite{nautsch2019privacy, treiber2019privacy} to the same problem.
Differently, Pathak et al. \cite{pathak2012matching}, Portêlo et al. \cite{sbeSV} and Jiménez et al. \cite{smh} have explored the applicability of distance-preserving hashing techniques to privacy-preserving ASV, while Mtibaa et al. have studied cancelable biometric schemes for ASV \cite{mtibaa2018cancelable, mtibaa2021privacy}. 

However, the above-mentioned works focus mainly on the security of the speaker templates or on how the verification step itself can be performed privately, sharing the assumption that the client locally extracts voice templates.
In contrast, we argue that this is extremely undesirable for service providers. Specifically, we argue that the model used to extract voice templates, or speaker embeddings, is one of the, if not the most valuable component in the speaker verification pipeline. This stems from the fact that speaker embedding extractors require large amounts of data and high levels of expertise to be developed. As such, by sharing this model, ASV service providers would relinquish control over their intellectual property, and consequently, lose the value it holds. Further, as noted by Das et al. \cite{das2020attackers} and Villalba et al. \cite{villalba2020xvectors}, having knowledge of the speaker embedding extractor model may allow attackers to craft adversarial examples that mislead the ASV system, raising security and robustness concerns.
For this reason, in this work we show how speaker embeddings can be extracted privately using Secure Multiparty Computation. Specifically, we focus on the private extraction of \textit{x-vector} speaker embeddings \cite{snyder2018xvectors}. This not only allows the protection of the speaker's voice, as it is never shared with the ASV provider, but also the protection of the speaker embedding extraction model. 
Moreover, even though we only consider the private extraction of \textit{x-vectors}, our implementation can be directly combined with some of the above-mentioned works for private speaker verification \cite{nautsch2019preserving, treiber2019privacy}, in order to produce a fully end-to-end private speaker verification system that protects both the speaker's voice and the vendor's model.

The remainder of this paper is organised as follows: in Section \ref{sec:smc} we provide the necessary background on Secure Multiparty Computation (SMC); Section \ref{sec:ppasv} specifies the setting and threat models assumed for our task; in Section \ref{sec:exp_setup} we describe the experimental setup, while in Section \ref{sec:results} we present and discuss the results obtained. Finally, Section \ref{sec:conclusions} presents our conclusions and topics for future work.

\section{Secure Multiparty Computation}
\label{sec:smc}
Secure Multiparty Computation (SMC) is an umbrella term for protocols designed to allow several parties to jointly and securely compute functions over their data, while keeping all inputs private.
SMC protocols are usually built over some form of Secret Sharing (e.g. Shamir's Secret Sharing \cite{shamir1979share}, GMW \cite{gmw}, BGW \cite{bgw}), or Garbled Circuits (e.g., Yao's GCs \cite{yao}, BMR \cite{beaver1990round}), and are often combined with cryptographic primitives like public-key encryption, symmetric encryption, Homomorphic Encryption (HE) or Oblivious Transfers (OTs) to perform specific functionalities, each lending different levels of security, computational and communication costs \cite{lindell2020secure}.
Our approach for the private extraction of  \textit{x-vectors} relies on two forms of secret sharing briefly described below. 
\vspace{-0.1cm}
\subsection{Secret Sharing}
\label{sec:smc:secret}
Secret sharing is a basic primitive for SMC protocols. It allows parties to represent and share their data with other parties in such a way that each of the parties participating in the computation will only have access to a random-looking \textit{share} (here denoted as $\share{\cdot}$) of the original value. 
Considering an additive secret sharing scheme in the $n$-party case, a value $x$ secret shared among several parties by a dealer, is defined as:
\begin{equation}
\label{eq:ss}
    x = \share{x}_1 + \share{x}_2 + ... + \share{x}_n,
\end{equation}
\noindent where $\share{x}_1$, ..., $\share{x}_n$ represent random-looking shares of $x$ held by each party, generated as $\share{x}_{n} = x - \sum_{i=1}^{n-1} s_i$, where each $s_i$ is chosen uniformly at random.
When a value is represented in this way, a single party is not able to reconstruct the \textit{secret} without the remaining parties.
Further, this representation allows for parties to interactively compute any operation over their secret data. 
For instance, due to the associative property of addition, adding two shared values simply amounts to each party adding the shares they hold that correspond to the two values, without requiring any communication with the other parties.

On the other hand, in the case of multiplications, additive secret sharing schemes require additional pre-computed random shared values called \textit{Multiplication Triples (MTs)} (or \textit{Beaver Triples} \cite{beaver1991efficient}). 
These values are shares of the form $\share{a}$, $\share{b}$ and $\share{c}$, where $\share{c}~=~\share{a} \times \share{b}$. 
To perform a multiplication between shared values $x$ and $y$, each party sets its shares to $\share{e}_i = \share{x}_i - \share{a}_i$ and $\share{f}_i = \share{y}_i - \share{b}_i$ and exchanges the results with the other parties, so that each party holds $e$ and $f$. The resulting share is given by \cite{aby}:

\begin{equation}
    \share{z}_i = i \cdot e \cdot f + f\cdot \share{a}_i + e\cdot \share{b}_i + \share{c}_i
\end{equation}

\noindent It can then be shown that by adding the $z_i$ we obtain $x \times y$.

The above works for any number of parties greater than or equal to two. However, for a number of parties strictly larger than two, it is possible to instantiate more efficient schemes. 
Replicated Secret Sharing (RSS) schemes \cite{araki2016high} are such an example. 
While in additive secret sharing, each party holds a single share per value in the computation, with RSS each party holds a set of shares per value.
Considering for instance the three party case and a shared value $y = \sum_{i=1}^3 \share{y}_i$, party $p_1$ will hold shares $\share{y}_1, \share{y}_2$, party $p_2$ will hold shares $\share{y}_2, \share{y}_3$ and party $p_3$ will hold shares $\share{y}_3, \share{y}_1$.
In this case, addition will work as before, and each party can simply perform the operation locally. Multiplication, on the other hand, may work differently. A possible implementation of the multiplication operation would be for each party to locally multiply the shares it holds for each of the secret shared values. In this way, party $p_1$ will obtain $z_1 = \share{x}_1 \share{y}_1 + \share{x}_1 \share{y}_2 + \share{x}_2 \share{y}_1$; party $p_2$, $z_2 = \share{x}_2 \share{y}_2 + \share{x}_2 \share{y}_3 + \share{x}_3 \share{y}_2$ and party $p_3$, $z_3 = \share{x}_3 \share{y}_3 + \share{x}_3 \share{y}_1 + \share{x}_1 \share{y}_3$. 
As above, it can be shown that adding the resulting shares will yield the correct result.
Still, at the end of the computation, each party only holds a single share of the value, and a \textit{re-sharing} protocol is required, so that each party holds the same set of shares as before \cite{wagh2021falcon}. 

The above-mentioned schemes are described with regard to arithmetic operations, but also hold for binary computations, with minor modifications \cite{gmw}. 
This is important, as performing operations in each of these domains may prove to be more efficient for different operations, or may even allow performing different functionalities.
Depending on the SMC protocol, the conversion between domains may take different forms. In some protocols, it is possible to convert between domains locally, with minimal interaction between parties \cite{dalskov2021fantastic, mpspdz}. However, other protocols may need to use pre-computed values that are shared in both domains such as \textit{daBits} \cite{rotaru2019marbled} or \textit{edaBits} \cite{escudero2020improved}.

The generation of \textit{Multiplication Triples}, \textit{daBits}, \textit{edaBits}, as well as other auxiliary secret shares -- called \textit{correlated randomness} -- requires the participation and interaction of all the parties involved in the computation. However, since the generation of these auxiliary shares is not dependent on input data, this step can be moved to what is called an \textit{offline}, or \textit{pre-processing} phase.
This phase can be performed at any time before the \textit{online} data-dependent phase.
Many protocols are hence designed to have the most expensive operations within the \textit{offline} phase, making the \textit{online} phase much more efficient.
\vspace{-0.1cm}
\subsection{Fixed-point numbers}
\label{sec:smc:fixed}
An important detail of SMC protocols is the fact that secret shared values are integers, whereas for most real-world applications, values are floating-point numbers. 
While floating-point representations exist within SMC, fixed-point representations are much more efficient. In this work, we adopt the fixed-point representation of \cite{mpspdz}, where a value $x$ is represented as $x = y \cdot 2^f$, where $y$ is an integer, and $f$ is the fixed precision.
While this approximation does not affect additions, for multiplications, one needs to first multiply the two integers, and then truncate by $f$. This can be implemented as a binary left shift operation \cite{catrina2010improved}, or via probabilistic truncation \cite{catrina2010secure, dalskov2020secure, escudero2020improved}.
\vspace{-0.2cm}
\subsection{Security}
The shared nature of SMC protocols demands making threat assumptions about the participating parties. The threat model of an SMC protocol is extremely important as it significantly affects its security, computational and communication performance, and thus, its range of applications.

The most common security (or threat) models include the \textit{semi-honest} adversary model (also called \textit{honest-but-curious} adversary or \textit{passive} security) and the \textit{malicious} adversary model (or \textit{active} security).
The \textit{honest-but-curious} model is the simplest model possible.
In this model, the adversaries are assumed to follow the established protocol, but are also assumed to pry into and record the data that is visible to them. In this way, no party will be able to obtain information other than that which it is allowed to, resulting in very efficient implementations. 
On the other hand, the \textit{malicious} model assumes that adversaries will attempt to thwart the protocol, demanding additional proof that each party is behaving correctly. This can be done in different ways, depending on the protocol and phase of the computation, through Zero Knowledge (ZK) proofs, cut-and-choose methods, Message Authentication Codes (MACs), among others \cite{spdz, spdz2k, mpspdz}. Although more secure, this model significantly increases the protocol's computational cost.

Besides the behaviour of individual parties, one can also define the security of the protocol in terms of whether a \textit{majority} of the parties will behave correctly or not -- \textit{honest majority} vs \textit{dishonest majority}, and whether a subset of parties might collaborate -- or \textit{collude} -- to obtain more information than they are allowed to. If a majority of parties are assumed to be honest, protocols that take into account \textit{malicious} behaviour can be made much more efficient \cite{furukawa2017high, dalskov2021fantastic}. 
On the other hand, the highest possible level of security is achieved by assuming malicious adversaries in a dishonest majority. However, this comes at very high computational and communication costs \cite{spdz, spdz2k}.
\vspace{-0.1cm}
\section{Privacy-preserving speaker embedding extraction}
\label{sec:ppasv}

We consider two parties to be involved in the extraction of speaker embeddings in the context of ASV: the \textit{client}, who wants to be able to access a given system and the \textit{ASV Vendor}, who provides the authentication system as a service.
If we were considering a full ASV system, we would also need to consider the \textit{ASV Controller}, the party who holds the set of speaker templates who are allowed to access the system. However, since in this work we only focus on the extraction of speaker embeddings, we will not take this party into account. Nonetheless, in some cases the \textit{vendor} and \textit{controller} may be same party. 

\begin{figure}
    \centering
    \resizebox{!}{0.4\columnwidth}{
    \includegraphics{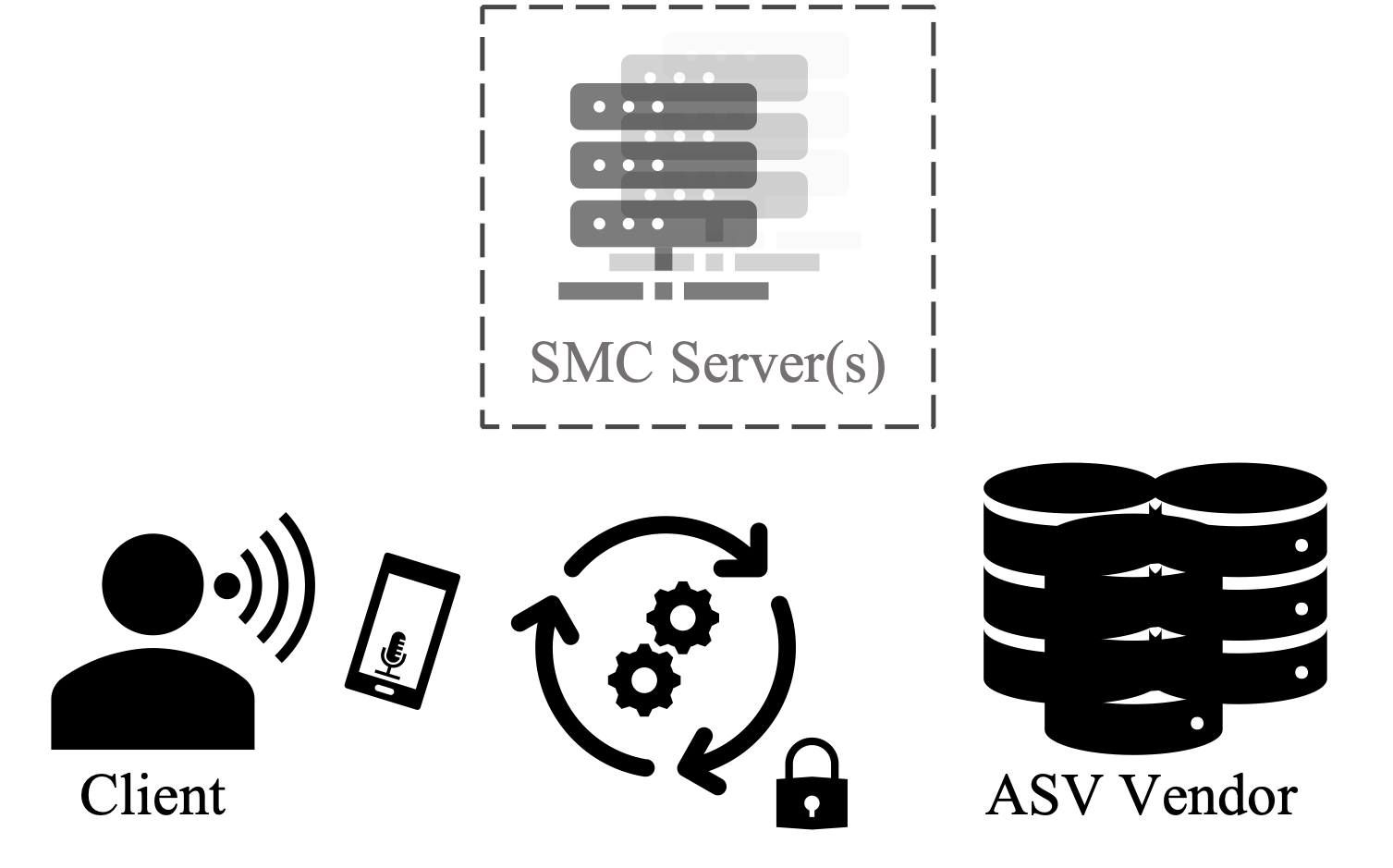}
    }
    \caption{Privacy-preserving extraction of speaker embeddings.}
    \label{fig:ppasv}
\end{figure}

\subsection{Threat models}
\label{sec:ppasv:threat}
Our goal is to have the \textit{vendor} and \textit{client} collaborate to privately extract a speaker embedding from a speech sample belonging to the \textit{client}, using an extraction model belonging to the \textit{vendor}. We consider that both the \textit{client} and \textit{vendor} are interested in protecting the privacy of their own data -- the \textit{client} for the sensitive nature of their speech data, and the \textit{vendor} due to the value and security of its model. To this end, we consider four scenarios.

In the first scenario, the \textit{vendor} and \textit{client} are the only parties involved in the private extraction of the speaker embedding, and are both assumed to be \textit{semi-honest}. This is the weakest security model, as either the \textit{vendor} or the \textit{client} might thwart the protocol to obtain information about the other's data. 

In the second scenario, we consider adding a trusted non-colluding SMC server to the computation. In a real world setting, this party would correspond to a company providing servers for SMC. Since such a company would need to rely on its reputation for its business, we argue that it would always follow protocol, and would never collude with any other party involved in the computation \cite{bogdanov2008sharemind}.  
By adding this trusted non-colluding server, and since the \textit{client} and \textit{vendor} have no incentive to collude -- the SMC server does not have data to be stolen -- this allows us to instantiate the honest majority 3-party RSS SMC protocol of Araki et al. \cite{araki2016high}. As discussed in Section \ref{sec:smc:secret}, RSS schemes are much more efficient than additive secret sharing protocols while keeping the same level of security \cite{araki2016high}.

For our third scenario, we consider adding a second trusted non-colluding SMC server. This allows us to instantiate a 4-party honest-majority RSS SMC protocol. Specifically, we can instantiate the 4-party protocol of Dalskov et al. \cite{dalskov2021fantastic}, which is secure against one malicious party. In this way, if either the \textit{client} or the \textit{vendor} behave maliciously, the protocol will abort. Since the \textit{client} and the \textit{vendor} will not collude, and the SMC servers are assumed to be trusted, this setting will be more secure than the previous. However, in this case the non-collusion assumption of the SMC server is much stronger.

In our fourth scenario we return to the 2-party setting, and assume that either the \textit{vendor} or the \textit{client} might behave maliciously. This is the setting with the highest level of security, however, it will also incur 
the highest computational and communication costs. A diagram of the aforementioned computational settings can be found in Figure \ref{fig:ppasv}.
\vspace{-0.1cm}
\subsection{Privacy-preserving \textit{x-vector} extraction using SMC}

Originally proposed by Snyder et al. \cite{snyder2018xvectors}, the \textit{x-vector} architecture is a neural network trained to discriminate between a large number of speakers. Within this context, \textit{x-vectors} correspond to latent representations extracted from an intermediate layer of the network.
This network is composed of three main blocks: the first block is a set of time-delay (TDNN) layers that operate at the frame level with a small temporal context. These layers work as 1D dilated convolutions, with a kernel size corresponding to the temporal context, which alternate with ReLU non-linearities; the second block, a statistical pooling layer, aggregates the information across the time dimension and outputs the per-feature mean and standard deviation for the entire speech segment; the third block is a set of fully connected layers, from which \textit{x-vector} embeddings are extracted after the network is trained for speaker classification. 

To implement this extractor with SMC, we need to take into account the type of operations required by each layer in the network.
1D dilated convolutional layers are linear transformations and can be implemented using either the basic arithmetic operations of the SMC protocol, or with specific protocols for inner product computations \cite{mpspdz}. ReLU activations require the computation of a comparison, which can be done through the secure comparison protocol of \cite{catrina2010secure}. The Statistical Pooling layer involves computing the mean and standard deviation of the input. To compute the standard deviation, we will need to compute a square-root, which can be done through the protocol of \cite{aly2019benchmarking}.

All the protocols mentioned above work for the fixed-point number representation of Section \ref{sec:smc:fixed}, making them directly compatible with the weights and inputs of neural networks, after these have also been converted to a fixed-point representation.
\vspace{-0.1cm}
\section{Experimental Setup}
\label{sec:exp_setup}

\begin{table*}[ht!]
 \caption{Results obtained for each protocol in terms of computational performance and communication cost.}
 \centering
\resizebox{\textwidth}{!}{
\begin{tabular}{@{}c|c|ccc|ccc@{}}
\toprule
\multirow{2}{*}{Protocol} & \multicolumn{1}{c|}{\multirow{2}{*}{\begin{tabular}[c]{@{}c@{}}Security\\ Model\end{tabular}}} & \multicolumn{3}{c|}{Time (s)} & \multicolumn{3}{c}{Communication (MB)} \\ \cmidrule(l){3-8} & \multicolumn{1}{c|}{} & Pre-processing & Online & Total & Pre-processing & Online & Total \\ \midrule
2-party Semi$_{2^k}$ \cite{spdz2k} & DM/SH  & 8,423.00.  $\pm$ 165.36$^{*}$   & 18.92  $\pm$ 0.22 & 8,441.93   $\pm$ 165.36 & 1,662,300.60$^{*}$    & 12,919.40 & 1,675,220.00  \\
3-party RSS \cite{araki2016high} & HM/SH  & 0.18       $\pm$ 0.20$^{*}$     & 10.68  $\pm$ 0.15 & 10.85      $\pm$ 0.14   & 15.04$^{*}$           & 118.02    & 133.06        \\
4-party RSS \cite{dalskov2021fantastic} & HM/Mal & 1.21       $\pm$ 0.29$^{*}$     & 16.76  $\pm$ 0.21 & 17.97      $\pm$ 0.21   & 27.14$^{*}$           & 333.16    & 360.30        \\
2-party SPDZ$_{2^k}$ \cite{spdz2k} & DM/Mal & 147,799.68 $\pm$ 1,016.82$^{*}$ & 126.32 $\pm$ 1.16 & 147,926.00 $\pm$ 1,016.82 & 21,870,489.60$^{*}$ & 27,810.40 & 21,898,300.00 \\ \bottomrule
\end{tabular}
}
\label{tab:results}
\end{table*}

\subsection{Corpora}
The Voxceleb corpus was used to train the \textit{x-vector} extractor and the Probabilistic Linear Discriminant Analysis (PLDA) model described below. This corpus includes recordings of 7,363 speakers of multiple ethnicities, accents, occupations and age groups. It is composed of short clips taken from interviews uploaded to YouTube \cite{nagrani2017voxceleb, voxceleb2}. The corpus is composed of two parts, \textit{VoxCeleb 1 and 2}, both subdivided into \textit{dev} and \textit{test}.

\subsection{Speaker embeddings}

For our experiments we used the pre-trained \textit{x-vector} model made available by SpeechBrain \cite{speechbrain}.
This model follows the architecture of \cite{snyder2018xvectors}. 
A description of the layers used for extraction can be found in Table \ref{tab:architecture}.
The model was trained using the \textit{dev} partitions of Voxceleb 1 and 2.
As a baseline reference for computational cost, extracting a single \textit{x-vector} from a 3-second long speech sample with this model, using a CPU,~takes~$\sim$0.03s.

\begin{table}[ht!]
\centering
\caption{\textit{x-vector} extractor architecture.}
\resizebox{!}{0.2\columnwidth}{
\begin{tabular}{@{}clrrrrr@{}}
\toprule
\# & Layer & Input & Output & Kernel & Dilation \\ \midrule
1  & TDNN 1       & 24    & 512  & 5  & 1 \\
2  & TDNN 2       & 512   & 512  & 3  & 2 \\
3  & TDNN 3       & 512   & 512  & 3  & 3 \\
4  & TDNN 4       & 512   & 512  & 1  & 1 \\
5  & TDNN 5       & 512   & 1500 & 1  & 1 \\
6  & Statistics Pooling & 1500  & 3000 & -  & - \\
7  & Linear             & 3000  & 512  & -  & - \\
\bottomrule
\end{tabular}
}
\label{tab:architecture}
\end{table}

\noindent A Probabilistic Linear Discriminant Analysis (PLDA) model was used to score pairs of \textit{x-vectors} when performing verification \cite{kenny2013plda}. 
The full pipeline achieves 3.2\% Equal Error Rate (EER) on the Voxceleb 1 test set (Cleaned) \cite{voxceleb2, speechbrain}.

\subsection{Privacy-preserving implementation}
The network described in the previous subsection was implemented using the MP-SPDZ library \cite{mpspdz}. 
We tested our implementation using four different protocols with different levels of security, as detailed in Section \ref{sec:ppasv:threat}. 
Specifically, we tested our implementation over the following protocols: the 2-party \textit{semi-honest} (SH) version of the SPDZ$_{2^k}$ scheme for dishonest majority (DM), denoted as Semi$_{2^k}$ \cite{spdz2k}; the 3-party RSS scheme described in \cite{araki2016high}, which provides \textit{semi-honest} security, in the \textit{honest majority} (HM) setting; the 4-party RSS scheme of \cite{dalskov2021fantastic}, which provides \textit{malicious} (Mal) security in the \textit{honest majority} setting against one corrupted party; and the 2-party \textit{malicious} version of the SPDZ$_{2^k}$ scheme \cite{spdz2k}.

For 3 and 4-party RSS and for the \textit{semi-honest} version of SPDZ$_{2^k}$, we used local share conversions \cite{dalskov2021fantastic} to improve efficiency. For 3 and 4-party RSS, we also used probabilistic truncation as proposed by \cite{dalskov2020secure, dalskov2021fantastic}, instead of regular truncation, to further improve efficiency.
Our experiments assume the default security parameters for each protocol, namely 40-bit security for 3 and 4-party RSS, and Semi$_{2^k}$, and 64-bit security for SPDZ$_{2^k}$. 
We used the library's fixed-point number representation adopting the default configuration of 16 bits for the decimal part, and 15 bits for the fractional part.
All tested protocols perform computations modulo $2^{k}$, where $k = 64$. Experiments were performed on a machine with 24 Intel(R) Xeon(R) CPU E5-2630 v2 @ 2.60GHz processors and 250GB of RAM.

\section{Results}
\label{sec:results}
Table \ref{tab:results} includes the results obtained for our experiments in terms of computational performance and communication cost. All results correspond to the extraction of a single \textit{x-vector} using a 3-second long speech sample. Online results for all protocols, and for full results for 3 and 4-party RSS, were obtained by averaging over 100 runs. Full results for Semi2k and SPDZ2k were computed over 10 runs, due to their high computational cost. Values denoted with $^{*}$ were estimated by computing the difference between the full protocol and the online phase.

Our results show that RSS schemes significantly outperform the \textit{semi-honest} and \textit{malicious} versions of SPDZ$_{2^k}$, both in terms of computational and communication performance. 
Further, since for the \textit{semi-honest} version of SPDZ$_{2^k}$, pre-processing takes $>$2h and $>$1TB of data, and since the pre-processing for the malicious version takes $>$41h and $>$20TB of data, it is clear that the private extraction of \textit{x-vectors} with these protocols, particularly for a high level of security, is currently infeasible. 
Contrarily, the results for the RSS schemes can be deemed feasible, particularly when considering the fact that no modifications were made to reduce the size of the \textit{x-vector} extraction network. 
When comparing the 3 and 4 party RSS protocols, while the 3-party \textit{semi-honest} version is more efficient in terms of computational cost and communication, we argue that the added security of the 4-party RSS protocol is a reasonable trade-off for the $\sim$7s and $\sim$230MB difference in the total computational and communication costs. Still, we need to consider that to implement the 4-party RSS protocol we need strong assumptions about the honest behaviour of the SMC servers.

Finally, our experiments showed that the SMC implementation yielded negligible degradation, with the average Mean Squared Error distance between 100 \textit{x-vectors} extracted with the original and SMC implementations being just $\sim1\%$ of the total magnitude of the vector.

\section{Conclusions}
\label{sec:conclusions}
In this work we have shown that it is possible to extract \textit{x-vector} speaker embeddings at a reasonable level of security and computational and communication costs, while protecting the privacy of both the \textit{client}'s data and the \textit{ASV Vendor}'s model, using SMC, particularly when deploying on 3 and 4-party RSS protocols.
This problem had been unexplored so far, as other privacy-preserving works for ASV assumed that speaker embeddings to be extracted by the client. This makes this work complementary to others in the literature and another step towards fully private ASV pipelines.

As future work, we consider that it would be important to explore ways to reduce the size of the \textit{x-vector} extraction network, to improve efficiency. Moreover, it would be interesting to also consider protocols following the \textit{covert} security model, wherein adversaries may have a malicious behaviour, but where there is a probability that in doing so, they may be discovered. 

\newpage
\bibliographystyle{IEEEtran}
\bibliography{mybib}

\begin{thebibliography}{10}
\providecommand{\url}[1]{#1}
\csname url@samestyle\endcsname
\providecommand{\newblock}{\relax}
\providecommand{\bibinfo}[2]{#2}
\providecommand{\BIBentrySTDinterwordspacing}{\spaceskip=0pt\relax}
\providecommand{\BIBentryALTinterwordstretchfactor}{4}
\providecommand{\BIBentryALTinterwordspacing}{\spaceskip=\fontdimen2\font plus
\BIBentryALTinterwordstretchfactor\fontdimen3\font minus
  \fontdimen4\font\relax}
\providecommand{\BIBforeignlanguage}[2]{{%
\expandafter\ifx\csname l@#1\endcsname\relax
\typeout{** WARNING: IEEEtran.bst: No hyphenation pattern has been}%
\typeout{** loaded for the language `#1'. Using the pattern for}%
\typeout{** the default language instead.}%
\else
\language=\csname l@#1\endcsname
\fi
#2}}
\providecommand{\BIBdecl}{\relax}
\BIBdecl

\bibitem{singh2019profiling}
R.~Singh, \emph{Profiling humans from their voice}.\hskip 1em plus 0.5em minus
  0.4em\relax Springer, 2019, {ISBN:} 978-981-13-8403-5.

\bibitem{nautsch2019preserving}
A.~Nautsch, A.~Jim{\'e}nez, A.~Treiber, J.~Kolberg, C.~Jasserand \emph{et~al.},
  ``Preserving privacy in speaker and speech characterisation,'' \emph{Computer
  Speech \& Language}, vol.~58, pp. 441--480, 2019.

\bibitem{gdpr}
{European Parliament and Council}, ``{Regulation (EU) 2016/679} of the
  {E}uropean {P}arliament and of the {C}ouncil of 27 {A}pril 2016 - {G}eneral
  {D}ata {P}rotection {R}egulation.'' 2016.

\bibitem{nautsch2019gdpr}
A.~Nautsch, C.~Jasserand, E.~Kindt, M.~Todisco, I.~Trancoso, and N.~Evans,
  ``The {GDPR} \& {S}peech data: Reflections of legal and technology
  communities, first steps towards a common understanding,'' \emph{arXiv
  preprint}, vol. 1907.03458, 2019.

\bibitem{pathak2013privacy}
M.~A. Pathak and B.~Raj, ``Privacy-{P}reserving {S}peaker {V}erification and
  {I}dentification {U}sing {G}aussian {M}ixture {M}odels,'' \emph{IEEE TASLP},
  vol.~21, no.~2, pp. 397--406, 2013.

\bibitem{portelo2014privacy}
J.~Port{\^e}lo, B.~Raj, A.~Abad, and I.~Trancoso, ``Privacy-preserving speaker
  verification using garbled gmms,'' in \emph{22nd (EUSIPCO)}.\hskip 1em plus
  0.5em minus 0.4em\relax IEEE, 2014, pp. 2070--2074.

\bibitem{nautsch2018homomorphic}
A.~Nautsch, S.~Isadskiy, J.~Kolberg, M.~Gomez-Barrero, and C.~Busch,
  ``{Homomorphic Encryption for Speaker Recognition: Protection of Biometric
  Templates and Vendor Model Parameters },'' in \emph{Proc. Odyssey}, 2018, pp.
  16--23.

\bibitem{nautsch2019privacy}
A.~Nautsch, J.~Patino, A.~Treiber \emph{et~al.}, ``{Privacy-Preserving Speaker
  Recognition with Cohort Score Normalisation},'' in \emph{Proc. Interspeech
  2019}, 2019, pp. 2868--2872.

\bibitem{treiber2019privacy}
A.~Treiber, A.~Nautsch, J.~Kolberg, T.~Schneider, and C.~Busch,
  ``Privacy-preserving {PLDA} speaker verification using outsourced secure
  computation,'' \emph{Speech Communication}, vol. 114, pp. 60--71, 2019.

\bibitem{pathak2012matching}
M.~A. Pathak and B.~Raj, ``Privacy-preserving speaker verification as password
  matching,'' in \emph{ICASSP}, 2012, pp. 1849--1852.

\bibitem{sbeSV}
J.~Port\^elo, A.~Abad, B.~Raj, and I.~Trancoso, ``Secure {B}inary {E}mbeddings
  of {F}ront-end {F}actor {A}nalysis for {P}rivacy {P}reserving {S}peaker
  {V}erification,'' in \emph{Interspeech}, 2013, pp. 2494--2498.

\bibitem{smh}
A.~Jim{\'e}nez, B.~Raj, J.~Port{\^e}lo, and I.~Trancoso, ``Secure {M}odular
  {H}ashing,'' in \emph{WIFS}.\hskip 1em plus 0.5em minus 0.4em\relax IEEE,
  2015, pp. 1--6.

\bibitem{mtibaa2018cancelable}
A.~Mtibaa, D.~Petrovska-Delacretaz, and A.~B.~Hamida, ``Cancelable speaker
  verification system based on binary gaussian mixtures,'' in \emph{4th ATSIP},
  2018, pp. 1--6.

\bibitem{mtibaa2021privacy}
A.~Mtibaa, D.~Petrovska-Delacr{\'e}taz, J.~Boudy, and A.~B.~Hamida,
  ``Privacy-preserving speaker verification system based on binary i-vectors,''
  \emph{IET Biometrics}, vol.~10, no.~3, pp. 233--245, 2021.

\bibitem{das2020attackers}
R.~K. Das, X.~Tian, T.~Kinnunen, and H.~Li, ``{The Attacker’s Perspective on
  Automatic Speaker Verification: An Overview},'' in \emph{Proc. Interspeech
  2020}, 2020, pp. 4213--4217.

\bibitem{villalba2020xvectors}
J.~Villalba, Y.~Zhang, and N.~Dehak, ``{x-Vectors Meet Adversarial Attacks:
  Benchmarking Adversarial Robustness in Speaker Verification},'' in
  \emph{Proc. Interspeech 2020}, 2020, pp. 4233--4237.

\bibitem{snyder2018xvectors}
D.~Snyder, D.~Garcia-Romero, G.~Sell, D.~Povey, and S.~Khudanpur, ``X-vectors:
  Robust {DNN} embeddings for speaker recognition,'' in \emph{Proc. ICASSP},
  Calgary, AB, Canada, April 2018.

\bibitem{shamir1979share}
A.~Shamir, ``How to share a secret,'' \emph{Communications of the ACM},
  vol.~22, no.~11, pp. 612--613, 1979.

\bibitem{gmw}
O.~Goldreich, S.~Micali, and A.~Wigderson, ``How to play any mental game,'' in
  \emph{Proc. 19th STOC}.\hskip 1em plus 0.5em minus 0.4em\relax ACM, 1987, pp.
  218--229.

\bibitem{bgw}
M.~Ben-Or, S.~Goldwasser, and A.~Wigderson, ``Completeness theorems for
  non-cryptographic fault-tolerant distributed computation,'' in \emph{Proc.
  20th STOC}.\hskip 1em plus 0.5em minus 0.4em\relax ACM, 1988, pp. 1--10.

\bibitem{yao}
A.~C. Yao, ``How to generate and exchange secrets,'' in \emph{27th SFCS}, 1986,
  pp. 162--167.

\bibitem{beaver1990round}
D.~Beaver, S.~Micali, and P.~Rogaway, ``The round complexity of secure
  protocols,'' in \emph{Proc. 22nd STOC}, 1990, pp. 503--513.

\bibitem{lindell2020secure}
Y.~Lindell, ``Secure multiparty computation ({MPC}).'' \emph{IACR Cryptology
  ePrint Archive}, vol. 2020, p. 300, 2020.

\bibitem{beaver1991efficient}
D.~Beaver, ``Efficient multiparty protocols using circuit randomization,'' in
  \emph{CRYPTO}.\hskip 1em plus 0.5em minus 0.4em\relax Springer, 1991, pp.
  420--432.

\bibitem{aby}
D.~Demmler, T.~Schneider, and M.~Zohner, ``{ABY} - a framework for efficient
  mixed-protocol secure two-party computation,'' in \emph{NDSS}, 2015.

\bibitem{araki2016high}
T.~Araki, J.~Furukawa, Y.~Lindell, A.~Nof, and K.~Ohara, ``High-throughput
  semi-honest secure three-party computation with an honest majority,'' in
  \emph{SIGSAC}.\hskip 1em plus 0.5em minus 0.4em\relax ACM, 2016, p.
  805–817.

\bibitem{wagh2021falcon}
S.~Wagh, S.~Tople, F.~Benhamouda, E.~Kushilevitz, P.~Mittal, and T.~Rabin,
  ``Falcon: Honest-majority maliciously secure framework for private deep
  learning,'' \emph{Proceedings on Privacy Enhancing Technologies}, vol.~1, pp.
  188--208, 2021.

\bibitem{dalskov2021fantastic}
A.~Dalskov, D.~Escudero, and M.~Keller, ``Fantastic four: Honest-majority
  four-party secure computation with malicious security,'' in \emph{30th
  {USENIX} Security Symposium}, 2021.

\bibitem{mpspdz}
M.~Keller, ``Mp-spdz: A versatile framework for multi-party computation,''
  \emph{Cryptology ePrint Archive}, vol. Report 2020/521, 2020.

\bibitem{rotaru2019marbled}
D.~Rotaru and T.~Wood, ``Marbled circuits: Mixing arithmetic and boolean
  circuits with active security,'' in \emph{International Conference on
  Cryptology in India}.\hskip 1em plus 0.5em minus 0.4em\relax Springer, 2019,
  pp. 227--249.

\bibitem{escudero2020improved}
D.~Escudero, S.~Ghosh, M.~Keller, R.~Rachuri, and P.~Scholl, ``Improved
  primitives for mpc over mixed arithmetic-binary circuits,'' in \emph{Annual
  International Cryptology Conference}.\hskip 1em plus 0.5em minus 0.4em\relax
  Springer, 2020, pp. 823--852.

\bibitem{catrina2010improved}
O.~Catrina and S.~d. Hoogh, ``Improved primitives for secure multiparty integer
  computation,'' in \emph{International Conference on Security and Cryptography
  for Networks}.\hskip 1em plus 0.5em minus 0.4em\relax Springer, 2010, pp.
  182--199.

\bibitem{catrina2010secure}
O.~Catrina and A.~Saxena, ``Secure computation with fixed-point numbers,'' in
  \emph{International Conference on Financial Cryptography and Data
  Security}.\hskip 1em plus 0.5em minus 0.4em\relax Springer, 2010, pp. 35--50.

\bibitem{dalskov2020secure}
A.~Dalskov, D.~Escudero, and M.~Keller, ``Secure evaluation of quantized neural
  networks,'' \emph{Proceedings on Privacy Enhancing Technologies}, vol.~4, pp.
  355--375, 2020.

\bibitem{spdz}
I.~Damg{\aa}rd, M.~Keller, E.~Larraia, V.~Pastro, P.~Scholl, and N.~P. Smart,
  ``Practical covertly secure mpc for dishonest majority--or: breaking the spdz
  limits,'' in \emph{European Symposium on Research in Computer
  Security}.\hskip 1em plus 0.5em minus 0.4em\relax Springer, 2013, pp. 1--18.

\bibitem{spdz2k}
R.~Cramer, I.~Damg{\aa}rd, D.~Escudero, P.~Scholl, and C.~Xing, ``Spd
  $\mathbb{Z}_{2^{k}}$: efficient mpc mod $2^{k}$ for dishonest majority,'' in
  \emph{Annual International Cryptology Conference}.\hskip 1em plus 0.5em minus
  0.4em\relax Springer, 2018, pp. 769--798.

\bibitem{furukawa2017high}
J.~Furukawa, Y.~Lindell, A.~Nof, and O.~Weinstein, ``High-throughput secure
  three-party computation for malicious adversaries and an honest majority,''
  in \emph{EUROCRYPT}.\hskip 1em plus 0.5em minus 0.4em\relax Springer, 2017,
  pp. 225--255.

\bibitem{bogdanov2008sharemind}
D.~Bogdanov, S.~Laur, and J.~Willemson, ``Sharemind: A framework for fast
  privacy-preserving computations,'' in \emph{European Symposium on Research in
  Computer Security}.\hskip 1em plus 0.5em minus 0.4em\relax Springer, 2008,
  pp. 192--206.

\bibitem{aly2019benchmarking}
A.~Aly and N.~P. Smart, ``Benchmarking privacy preserving scientific
  operations,'' in \emph{International Conference on Applied Cryptography and
  Network Security}.\hskip 1em plus 0.5em minus 0.4em\relax Springer, 2019, pp.
  509--529.

\bibitem{nagrani2017voxceleb}
A.~Nagrani, J.~S. Chung, and A.~Zisserman, ``Voxceleb: A large-scale speaker
  identification dataset,'' \emph{Interspeech 2017}, Aug 2017.

\bibitem{voxceleb2}
J.~S. Chung, A.~Nagrani, and A.~Zisserman, ``Voxceleb2: Deep speaker
  recognition,'' \emph{Interspeech 2018}, Sep 2018.

\bibitem{speechbrain}
M.~Ravanelli, T.~Parcollet, P.~Plantinga, A.~Rouhe, S.~Cornell \emph{et~al.},
  ``{SpeechBrain}: A general-purpose speech toolkit,'' 2021, arXiv:2106.04624.

\bibitem{kenny2013plda}
P.~Kenny, T.~Stafylakis, P.~Ouellet, M.~J. Alam, and P.~Dumouchel, ``{PLDA} for
  speaker verification with utterances of arbitrary duration,'' in
  \emph{ICASSP}.\hskip 1em plus 0.5em minus 0.4em\relax IEEE, 2013, pp.
  7649--7653.

\end{thebibliography}

\end{document}